\documentclass[11pt]{article}
\usepackage[english]{babel}
\usepackage[a4paper,top=2cm,bottom=2cm,left=2.5cm,right=2.5cm,marginparwidth=1.75cm]{geometry}
\usepackage{setspace}
\usepackage{amsmath}
\usepackage{graphicx}
\usepackage[utf8]{inputenc}
\usepackage[T1]{fontenc}
\usepackage{amsmath,amssymb}
\usepackage[numbers]{natbib}
\usepackage{titlesec}
\usepackage{caption}
\usepackage{multicol}
\usepackage{ragged2e} 
\usepackage[dvipsnames]{xcolor} 
\usepackage{tcolorbox} 
\usepackage[colorlinks=true, allcolors=blue]{hyperref}
\usepackage[affil-sl]{authblk}

\title{Why the Northern Hemisphere Needs a 30--40 m Telescope\\ and the Science at Stake.\\ How do Planetary Systems Form?}
\author[1]{I. Mendigutía\thanks{\texttt{imendigutia@cab.inta-csic.es}}}
\author[1]{N. Huélamo\thanks{\texttt{nhuelamo@cab.inta-csic.es}}}
\author[1]{I. Jiménez-Serra}
\author[2,3]{E. Villaver}
\author[1]{O. Balsalobre-Ruza}
\author[1]{D. Barrado}
\author[4]{M. Benisty}
\author[5]{A. Boccaletti}
\author[6,7]{H. Bouy}
\author[4]{G. Chauvin}
\author[8]{G. Cugno}
\author[9]{R. Fedriani}
\author[9]{M. Fernández}
\author[1]{A. Fuente}
\author[10]{S. Haffert}
\author[11,12]{M. Kama}
\author[1]{J. Lillo-Box}
\author[13]{G. Meeus}
\author[14]{N. Miret-Roig}
\author[1]{B. Montesinos}
\author[9]{M. Osorio}
\author[15,16]{R.D. Oudmaijer}
\author[9]{A.F. Placinta-Mitrea}
\author[17]{D. Pollacco}
\author[18]{I. Rebollido}
\author[19]{M. Reggiani}
\author[20]{A. Ribas}
\author[21]{P. Rivière-Marichalar}
\author[22]{A. Sicilia-Aguilar}
\author[23]{C. Toci}
\author[4]{R. van Boekel}
\author[10]{N. van der Marel}
\author[24]{M. Vioque}
\author[25]{E. Whelan}
\author[26]{A. Zurlo}

\affil[1]{Centro de Astrobiología (CAB), CSIC-INTA, Camino Bajo del Castillo s/n, 28692, Villanueva de la Cañada, Madrid, Spain}
\affil[2]{Instituto de Astrofísica de Canarias, Vía Láctea s/n, 38200 La Laguna, Tenerife, Spain}
\affil[3]{Departamento de Astrofísica, Universidad de La Laguna, Astrofísico Francisco Sanchez s/n, 38206 La Laguna, Tenerife, Spain}
\affil[4]{Max-Planck Institute for Astronomy (MPIA), Königstuhl 17, 69117 Heidelberg, Germany}
\affil[5]{LESIA, Observatoire de Paris, Université PSL, CNRS, Sorbonne Université, Université de Paris, 5 place Jules Janssen, 92195 Meudon, France}
\affil[6]{Laboratoire d’astrophysique de Bordeaux, Univ. Bordeaux, CNRS, B18N, allée Geoffroy Saint-Hilaire, 33615 Pessac, France}
\affil[7]{Institut universitaire de France (IUF), 1 rue Descartes, 75231 Paris Cedex 05, France}
\affil[8]{Department of Astrophysics, University of Zurich, Winterthurerstrasse 190, 8057 Zurich, Switzerland}
\affil[9]{Instituto de Astrofísica de Andalucía, CSIC, Glorieta de la Astronomía, 18008 Granada, Spain}
\affil[10]{Leiden Observatory, Leiden University, Leiden, The Netherlands}
\affil[11]{Department of Physics and Astronomy, University College London, Gower Street, London, WC1E 6BT, UK}
\affil[12]{Tartu Observatory, University of Tartu, Observatooriumi 1, T\~{o}ravere, 61602, Estonia}
\affil[13]{Departamento Física Teórica, Facultad de Ciencias, Universidad Autónoma de Madrid, Campus de Cantoblanco, Carretera Colmenar s/n - km 15, 28049, Madrid, Spain}
\affil[14]{Institut de Ciències del Cosmos (ICCUB), Univ. de Barcelona (UB), Martí i Franquès, 1, 08028 Barcelona, Spain}
\affil[15]{Royal Observatory of Belgium, Ringlaan 3, 1180 Brussels, Belgium}
\affil[16]{School of Physics and Astronomy, University of Leeds, Leeds LS2 9JT, UK}
\affil[17]{Department of Physics, University of Warwick, Gibbet Hill Road, Coventry CV4 7AL, UK}
\affil[18]{European Space Agency (ESA), European Space Astronomy Centre (ESAC), Camino Bajo del Castillo s/n, 28692 Villanueva de la Cañada, Madrid, Spain}
\affil[19]{KU Leuven, Institute of Astronomy, Celestijnenlaan 200D, bus 2401, 3001 Leuven, Belgium}
\affil[20]{Astronomy Unit, Department of Physics and Astronomy, Queen Mary University of London, London, UK}
\affil[21]{Observatorio Astronómico Nacional (OAN, IGN), Calle Alfonso XII 3, 28014 Madrid, Spain}
\affil[22]{SUPA, School of Science and Engineering, University of Dundee, Nethergate, DD1 4HN, Dundee, UK}
\affil[23]{Departamento de Física Aplicada III, Universidad de Sevilla, Camino descubrimientos s/n, 41092, Sevilla, Spain}
\affil[24]{European Southern Observatory, Karl-Schwarzschild-Str. 2, 85748 Garching bei München, Germany}
\affil[25]{Department of Physics, Maynooth University, Maynooth, Co.Kildare, Ireland}
\affil[26]{Instituto de Estudios Astrofísicos, Facultad de Ingeniería y Ciencias, 3717 Universidad Diego Portales, Av. Ejército Libertador 441, Santiago, 3718, Chile}

\begin{document}
\maketitle
\newpage
\begin{tcolorbox}[colback=RoyalBlue!5!white,colframe=black!75!black, width=\textwidth]
\justifying
{\em  \noindent Current facilities have revealed the diversity of exoplanets around mature stars and the complex structures of protoplanetary disks around young stars, yet we lack the crucial observational link between them: a statistically meaningful census of protoplanets caught in the act of formation. Such a breakthrough requires a 30-40 m telescope that complements the ELT by covering the Northern hemisphere. That is key to obtaining diffraction-limited imaging of protoplanets and disks across the entire sky, enabling robust demographics, leveraging synergies with other observatories covering the North, and ensuring that Europe remains at the forefront of the planet-formation revolution in the coming decades.}
\end{tcolorbox}

\section{Scientific context}
\begin{figure}[h!]
    \centering
    \includegraphics[width=1\textwidth]{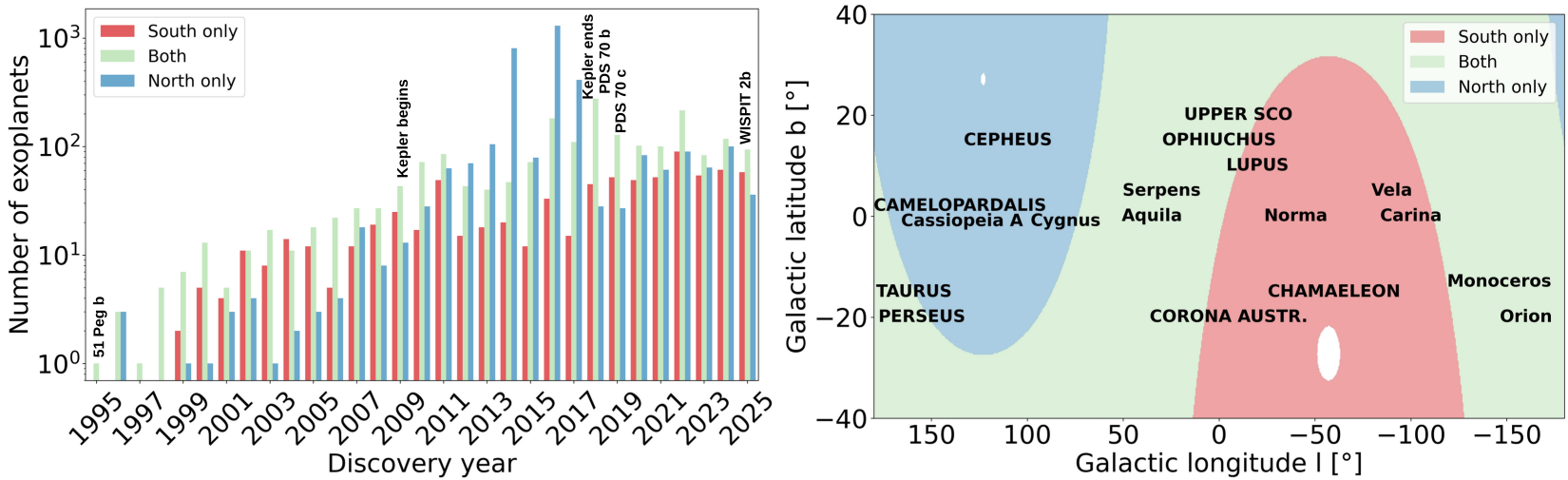}
    \caption{\small \textbf{(Left)} Exoplanet discoveries through the years (from the Encyclopaedia of exoplanetary systems; \href{https://exoplanet.eu}{https://exoplanet.eu/}). \textbf{(Right)} Galactic coordinates of the main star forming regions. The closest ones partially or totally located at distances $\leq$ 300 pc are indicated with capital letters. \textbf{(Both)} The colour code indicates whether the sources can be observed only from the South (declination $\delta$ $\lesssim$ -31$^{\circ}$), only from the North ($\delta$ $\gtrsim$ 35$^{\circ}$), or from both hemispheres. The Cerro Armazones and La Palma observatories are taken as reference, assuming that a source cannot be observed for elevations $<$ 30$^{\circ}$. Roughly half of the exoplanets and protoplanetary disks can be better or only covered from the North.}
    \label{fig:exoplanet_and_SFR_coverage}
\end{figure}

Thousands of exoplanets have been discovered since a planet around a star other than the Sun was first detected 30 years ago by the Nobel laureates \citep{Mayor95}. Instrumental developments and telescopes covering both hemispheres (Fig. \ref{fig:exoplanet_and_SFR_coverage}, left) have revealed the extraordinary abundance and diversity of the exoplanet zoo, which has changed our vision of Earth's place in the Universe. Nevertheless, almost all known exoplanets orbit around mature stars mostly in the main sequence (MS) phase. Consequently, our understanding of planet formation is incomplete as it is empirically based on already formed planets on the one hand, and the gas and dust properties of protoplanetary disks - the sites of planet formation - on the other. In particular, how and when dust grains and pebbles grow into planetesimals and planets, or the exact mechanisms driving early orbital migration, remain poorly understood \citep{Dra23}. The heterogeneous characteristics in terms of, e.g., initial structure, composition, and lifetime of protoplanetary disks with rings, gaps, spirals, and arcs/crescents, also present significant unknowns whose specific influence on planetary system outcomes is not yet clear \citep{Manara23,Bae23}. In short, we are still far from establishing a clear connection between the widely diverse protoplanetary disk's properties and the observed exoplanet populations. 

The key missing piece to bridge the previous gap is establishing a solid observational census of protoplanets forming in disks around young stars. Only recently we have started to detect this type of forming planets (Table \ref{table:pps}). Thus, the detection and characterization of protoplanets in disks is an emerging field that is taking off now and will represent a major revolution during the coming decades. The Extremely Large Telescope (ELT) will undoubtedly revolutionize protoplanet studies even with its first-light instruments \citep[e.g.][]{Chen22,Oberg23}, owing to its exquisite sensitivity and angular resolution reaching the mid-IR regime and capable of resolving the finest structures in protoplanetary disks. However, the low latitude of the Cerro Armazones observatory will inevitably leave half of the most relevant targets poorly covered or inaccessible. This white paper argues that this is not a trivial geographical issue, and that a 30-40 m telescope is also necessary in the Northern hemisphere for Europe to stay at the forefront of the planet formation revolution.

\begin{table}[h!]
\centering
\renewcommand\arraystretch{1}
\renewcommand\tabcolsep{5pt}
\caption{Representative subsample of confirmed and main candidate protoplanets}
\label{table:pps}
\centering
\begin{tabular}{l c l l l c}
\hline\hline
Name & Observability & Host & Discovery Technique & Status &  References \\
\hline\hline
PDS 70 b, c  & S      & T Tauri & AO spectro-imaging & Confirmed & \citep{Keppler2018,Haffert2019}\\
WISPIT 2b & S \& N & T Tauri & AO Imaging & Confirmed & \citep{vanCape2025}\\
IRAS 04125+2902 b & N  & T Tauri & Transit & Candidate & \citep{Barber2024}\\
AB Aur b          & N  & Herbig & AO Imaging & Candidate & \citep{Currie22}\\
MWC\,758 c        & N  & Herbig & AO Imaging & Candidate & \citep{Wagner2023}\\
HD169142 b        & S & Herbig & AO Imaging & Candidate & \citep{Hammond2023} \\
HD163296 b, c, d & S & Herbig  & Velocity Kinks & Candidates & \citep{Pinte2018,Teague2018,Izquierdo2022}\\
\hline
\hline
\end{tabular}
\begin{minipage}{16cm}
\small
\textbf{Notes:}  The "Observability" column refers to the hemisphere from which adaptive optics performance is best suited, meaning that the source has an elevation $\gtrsim$ 40$^{\circ}$ from Cerro Armazones (S) or La Palma (N). The last column indicates the discovery paper. \\
\end{minipage}
\end{table}   

\section{The Northern Opportunity: unique sky coverage}

Table \ref{table:pps} serves as a guide for future searches for protoplanets. First, most discoveries have required the use of high-angular resolution techniques. This implies that adaptive optics (AO) systems are essential to reach diffraction-limited imaging at optical/IR wavelengths, and long baseline interferometry (e.g. ALMA for HD 163296) is required at (sub-)mm wavelengths. This observational approach is not expected to change significantly, at least in the short term. Indeed, the classical radial velocity and transit methods valid for MS stars are not generally applicable to pre-MS stars because of the presence of variable accretion disks (IRAS 04125+2902 is an exception due to its particular star-disk-planet configuration). Second, although the most common young stellar objects are the low-mass "T Tauri" stars, many protoplanet candidates have more massive hosts. This is related to current limitations of high-angular resolution techniques and the fact that disks are generally larger and better resolved for less numerous, intermediate-mass "Herbig" stars. Similarly, future surveys should also consider relatively uncommon hosts, for which observing from both hemispheres is necessary for the best possible statistical coverage. Last but not least, although the number of known protoplanets is expected to significantly increase in the coming decades, each target will have the maximum scientific priority, and roughly half will require a Northern observatory to carry out high-quality AO-assisted observations. 

The angular resolution achievable by an AO-assisted, 35 m telescope is $\sim$ 4 - 9 mas in H$\alpha$ -a main tracer of ongoing planet formation- and the near-IR, which translates into a spatial resolution $\lesssim$ 1 - 3 au for distances closer than $\sim$ 300 pc. This high angular resolution is thus fundamental for detecting protoplanets in disks around young stars, which are mostly located in star-forming regions (SFRs). Future surveys will prioritise the nearest SFRs (indicated with capital letters in Fig. \ref{fig:exoplanet_and_SFR_coverage}, right), given that protoplanetary and circumplanetary disks, protoplanets, and their interactions can be observed in these regions at unprecedented levels of detail.

Taurus stands out as one of the most iconic SFRs. Not only does Taurus contain one of the richest collections of young sources with protoplanetary disks (including the prototypical "T-Tauri" star), but it is among the closest regions ($\sim$ 140 pc). However, because of its latitude, the ELT will observe Taurus only at low elevation. This will have a direct impact on the AO performance, which is severely degraded for elevations $\lesssim$ 40$^{\circ}$, causing protoplanets to become undetectable in the unmitigated glow of their host star. In addition, the thermal background will also be very high, affecting the sensitivity at mid-IR wavelengths, a key observing range to detect protoplanets and their circumplanetary disks \citep{Chen22}.

The large population of young stars in Taurus, along with those in the northern SFRs Cepheus and Perseus, represents roughly half of the most informative planet-forming laboratories in the solar neighbourhood. Observing them with a 30-40 m telescope in the North is essential to probe a wide range of physical conditions and build solid observational demographics of protoplanets in disks. By observing the most relevant SFRs in both hemispheres we will probe different evolutionary stages and environments that have an effect on planet formation, e.g. in terms of stellar and gas densities, external UV radiation or chemical composition \citep[e.g.][]{Parker20,Oberg21}. Moreover, a 30-40 m telescope in the North will give us access to other sources for which the samples are too scarce to be covered from a single hemisphere, probing not only Herbig stars, but also faint brown dwarfs and free-floating planets, or bright enough Massive Young Stellar Objects. 

Finally, it is a major advantage that many young stars, either in SFRs or isolated \citep{Delfini25}, 
are accessible from both hemispheres. Variability is not only a defining characteristic of young stars with circumstellar disks, but also of protoplanets \citep[e.g.][]{Zhou25}. Monitoring them over different timescales is essential to constrain protoplanet orbits and to access the temporal behaviour that encodes the physics of protoplanet assembly and the properties of their forming atmospheres.

\section{The Northern Opportunity: synergy with other facilities}
A 30-40 m telescope in the North would complement the capabilities of current and future space (e.g. JWST, Gaia, HWO, PLATO, Ariel and LIFE) and ground-based (e.g. NOEMA and next-generation VLA) facilities covering that hemisphere. Concerning the detection and characterization of protoplanets, here we emphasize the synergies with two major observatories. 

First, ngVLA \citep{Murphy18} will have superb sensitivity and resolution, and a larger coverage of the Northern sky compared to ALMA or SKA. The ngVLA will cover the frequency range between 1.2 to 116 GHz, probing the critical grain-growth barrier between millimetre to centimetre dust sizes, essential to understand pebble and planet formation. The maximum angular resolution achieved by the ngVLA will be of 0.2 mas at 30 GHz (1 cm), which will enable the imaging of the innermost regions in protoplanetary disks down to angular resolutions of 0.03 au at 140 pc, probing well into the rocky planet formation region in disks around Solar and M-type stars. The superb sensitivity of the ngVLA will measure the planetary initial mass function down to 5-10 Earth masses. Therefore, the synergy between a 30-40 m telescope and ngVLA will probe in a continuous manner the whole population of protoplanets from 5-10 Earth masses to gas giants. 

Particularly relevant is also the synergy with Gaia, which will provide between 7500 and 120,000 new planet candidates during DR4 (expected for December 2026) and DR5 ($\sim$ 2030) \citep{Lammers25}. This is a factor 10-100 larger than the currently known exoplanet population. Among the previous exoplanet candidates, a non-negligible fraction will be located in protoplanetary disks around young stars, given that Gaia astrometry serves to identify (sub-)stellar companion candidates also in such type of systems \citep{Ribas25,Vioque25}. The potential launch of GaiaNIR in 2040 (\href{https://www.gaianir.org/}{https://www.gaianir.org/}) could significantly expand the number of protoplanet candidates by providing 10-20 times better proper motions, including young embedded sources not covered by Gaia. Roughly half of all Gaia(NIR) candidates will be preferentially accessible from the North, and a 30-40 m telescope will be the best facility capable of confirming and characterizing them.

\section{Concluding remarks}
This white paper is one of a coordinated series demonstrating the scientific need for establishing a 30-40 m telescope in the Northern hemisphere. Here we have focused on protoplanets —objects that, like mature exoplanets, are exquisitely valuable. Each one offers a unique and irreplaceable view of planet formation in action. Missing those located in the North would mean losing the opportunity to fully understand how the variety of planetary systems is built.

Crucially, the core capabilities required for this science are already being developed through the ELT, which is a benchmark for future technological developments. What is required is committing those capabilities to a site that provides access to all the main planet-forming laboratories and to the synergies offered by major facilities covering the North. A Northern 30-40 m telescope would therefore not duplicate the ELT but complement it, filling the observational gap that currently limits our ability to trace the origins of planetary systems.

\begin{multicols}{2}
\renewcommand{\bibfont}{\footnotesize}
\bibliographystyle{unsrtnat}
\bibliography{references.bib}

@ARTICLE{Mayor95,
       author = {{Mayor} \& {Queloz}},
       journal = {Nature},
         year = 1995,
       volume = {378},
        pages = {355},
       adsurl = {https://ui.adsabs.harvard.edu/abs/1995Natur.378..355M}
}

@ARTICLE{Chen22,
       author = {{Chen} \& {Szul{\'a}gyi}},
      journal = {MNRAS},
         year = 2022,
       volume = {516},
        pages = {506},
archivePrefix = {arXiv},
       eprint = {2112.12821},
 primaryClass = {astro-ph.EP},
       adsurl = {https://ui.adsabs.harvard.edu/abs/2022MNRAS.516..506C}
}

@ARTICLE{Wagner2023,
       author = {{Wagner} et al.},
      journal = {Nat-Ast},
         year = 2023,
       volume = {7},
        pages = {1208},
archivePrefix = {arXiv},
       eprint = {2307.04021},
 primaryClass = {astro-ph.EP},
       adsurl = {https://ui.adsabs.harvard.edu/abs/2023NatAs...7.1208W}
}

@ARTICLE{Hammond2023,
       author = {{Hammond} et al.},
      journal = {MNRAS},
         year = 2023,
       volume = {522},
        pages = {L51},
archivePrefix = {arXiv},
       eprint = {2302.11302},
 primaryClass = {astro-ph.EP},
       adsurl = {https://ui.adsabs.harvard.edu/abs/2023MNRAS.522L..51H}
}

@ARTICLE{Izquierdo2022,
       author = {{Izquierdo} et al.},
      journal = {ApJ},
         year = 2022,
       volume = {928},
          eid = {2},
        pages = {2},
archivePrefix = {arXiv},
       eprint = {2111.06367},
 primaryClass = {astro-ph.EP},
       adsurl = {https://ui.adsabs.harvard.edu/abs/2022ApJ...928....2I}
}

@ARTICLE{Pinte2018,
       author = {{Pinte} et al.},
      journal = {ApJL},
         year = 2018,
       volume = {860},
          eid = {L13},
        pages = {L13},
archivePrefix = {arXiv},
       eprint = {1805.10293},
 primaryClass = {astro-ph.SR},
       adsurl = {https://ui.adsabs.harvard.edu/abs/2018ApJ...860L..13P}
}

@ARTICLE{Teague2018,
       author = {{Teague} et al.},
      journal = {ApJL},
         year = 2018,
       volume = {860},
          eid = {L12},
        pages = {L12},
archivePrefix = {arXiv},
       eprint = {1805.10290},
 primaryClass = {astro-ph.EP},
       adsurl = {https://ui.adsabs.harvard.edu/abs/2018ApJ...860L..12T}
}

@ARTICLE{Keppler2018,
       author = {{Keppler} et al.},
      journal = {A\&A},
         year = 2018,
       volume = {617},
          eid = {A44},
        pages = {A44},
archivePrefix = {arXiv},
       eprint = {1806.11568},
 primaryClass = {astro-ph.EP},
       adsurl = {https://ui.adsabs.harvard.edu/abs/2018A&A...617A..44K}
}

@ARTICLE{Haffert2019,
       author = {{Haffert} et al.},
      journal = {Nat-Ast},
         year = 2019,
       volume = {3},
        pages = {749},
archivePrefix = {arXiv},
       eprint = {1906.01486},
 primaryClass = {astro-ph.EP},
       adsurl = {https://ui.adsabs.harvard.edu/abs/2019NatAs...3..749H}
}

@ARTICLE{Barber2024,
       author = {{Barber} et al.},
      journal = {Nature},
         year = 2024,
       volume = {635},
        pages = {574},
archivePrefix = {arXiv},
       eprint = {2411.18683},
 primaryClass = {astro-ph.EP},
       adsurl = {https://ui.adsabs.harvard.edu/abs/2024Natur.635..574B}
}

@ARTICLE{vanCape2025,
       author = {{van Capelleveen} et al.},
      journal = {ApJL},
         year = 2025,
       volume = {990},
          eid = {L8},
        pages = {L8},
archivePrefix = {arXiv},
       eprint = {2508.19053},
 primaryClass = {astro-ph.EP},
       adsurl = {https://ui.adsabs.harvard.edu/abs/2025ApJ...990L...8V}
}

@ARTICLE{Currie22,
       author = {{Currie} et al.},
      journal = {Nat-Ast},
         year = 2022,
       volume = {6},
        pages = {751},
archivePrefix = {arXiv},
       eprint = {2204.00633},
 primaryClass = {astro-ph.EP},
       adsurl = {https://ui.adsabs.harvard.edu/abs/2022NatAs...6..751C}
}

@article{Zhou25,
year = {2025},
publisher = {The American Astronomical Society},
volume = {980},
pages = {L39},
author = {{Zhou} et al.},
journal = {ApJL},
}

@INPROCEEDINGS{Bae23,
       author = {{Bae} et al.},
    booktitle = {Protostars and Planets VII},
         year = 2023,
       series = {ASPC},
       volume = {534},
        pages = {423},
 primaryClass = {astro-ph.EP},
       adsurl = {https://ui.adsabs.harvard.edu/abs/2023ASPC..534..423B}
}

@INPROCEEDINGS{Manara23,
       author = {{Manara} et al.},
    booktitle = {Protostars and Planets VII},
         year = 2023,
       series = {ASPC},
       volume = {534},
        pages = {539},
 primaryClass = {astro-ph.SR},
       adsurl = {https://ui.adsabs.harvard.edu/abs/2023ASPC..534..539M}
}

@INPROCEEDINGS{Dra23,
       author = {{Dr{\k{a}}{\.z}kowska} et al.},
    booktitle = {Protostars and Planets VII},
         year = 2023,
       series = {ASPC},
       volume = {534},
        pages = {717},
 primaryClass = {astro-ph.EP},
       adsurl = {https://ui.adsabs.harvard.edu/abs/2023ASPC..534..717D}
}

@ARTICLE{Parker20,
       author = {{Parker}.},
      journal = {Royal Society Open Science},
         year = 2020,
       volume = {7},
       number = {11},
archivePrefix = {arXiv},
       eprint = {2007.07890},
 primaryClass = {astro-ph.EP},
       adsurl = {https://ui.adsabs.harvard.edu/abs/2020RSOS....701271P}
}

@ARTICLE{Oberg21,
       author = {{{\"O}berg} \& {Bergin}},
      journal = {Physics Reports},
         year = 2021,
       volume = {893},
        pages = {1},
archivePrefix = {arXiv},
       eprint = {2010.03529},
 primaryClass = {astro-ph.EP},
       adsurl = {https://ui.adsabs.harvard.edu/abs/2021PhR...893....1O}
}

@ARTICLE{Murphy18,
       author = {{Murphy} et al.},
       journal = {ASPC},
         year = 2018,
      volume = {517},
        pages = {3},
archivePrefix = {arXiv},
       eprint = {1810.07524},
 primaryClass = {astro-ph.IM},
       adsurl = {https://ui.adsabs.harvard.edu/abs/2018ASPC..517....3M}
}

@ARTICLE{Vioque25,
author = {{Vioque} et al.},
 journal = {A\&A, accepted},
volume = {arXiv:2512.00157},
}

@ARTICLE{Lammers25,
        author = {{Lammers} \& {Winn}},
        journal = {AJ, accepted},
        volume = {arXiv:2511.04673},
}

@ARTICLE{Oberg23,
       author = {{{\"O}berg} et al.},
      journal = {A\&A},
         year = 2023,
       volume = {670},
          eid = {A74},
        pages = {A74},
archivePrefix = {arXiv},
       eprint = {2212.03007},
 primaryClass = {astro-ph.EP},
       adsurl = {https://ui.adsabs.harvard.edu/abs/2023A&A...670A..74O}
}

@ARTICLE{Delfini25,
       author = {{Delfini} et al.},
      journal = {A\&A},
         year = 2025,
       volume = {699},
          eid = {A145},
        pages = {A145},
archivePrefix = {arXiv},
       eprint = {2505.04699},
 primaryClass = {astro-ph.SR},
       adsurl = {https://ui.adsabs.harvard.edu/abs/2025A&A...699A.145D}
}

@ARTICLE{Ribas25,
       author = {{Ribas} et al.},
      journal = {Nat-Ast},
         year = 2025,
       volume = {9},
        pages = {1176},
 primaryClass = {astro-ph.EP},
       adsurl = {https://ui.adsabs.harvard.edu/abs/2025NatAs...9.1176R}
}
\end{multicols}

\end{document}